\documentclass[12pt]{iopart}
\usepackage{graphicx}
\usepackage{iopams}
\usepackage{amssymb}
\usepackage{multirow}
\usepackage{float}

\newcommand{\ped}[1]{\ensuremath{_{\rm #1}}}

\begin{document}
\title{Eliashberg theory of a multiband non-phononic spin glass superconductor}
\author{G.A. Ummarino$^{1,2}$}
\ead{giovanni.ummarino@polito.it}
\address{$^1$ Istituto di Ingegneria e Fisica dei Materiali, Dipartimento di Scienza Applicata e Tecnologia, Politecnico di
Torino, Corso Duca degli Abruzzi 24, 10129 Torino, Italy}
\address{$^2$National Research Nuclear University MEPhI (Moscow Engineering Physics Institute), Kashirskoe shosse 31, Moscow 115409, Russia}

\begin{abstract}
I solved the Eliashberg equations for multiband non-phononic $s\pm$ wave spin-glass superconductor
and I calculated the temperature dependence of gaps and superfluid density, revealing unusual behaviors such as non monotonic temperature dependence and reentrant superconductivity.
The phase diagram, for particular values of input parameters that could describe the iron pnictide $EuFe_{2}(As_{1-x}P_{x})_{2}$, is still more complex with two different ranges of temperature where the superconductivity appears.
\end{abstract}
\maketitle
\section{Introduction}
The discovery of the new iron-based superconductors family based on $EuFe_{2}As_{2}$ \cite{Eu1,Eu2,Eu3,Eu4,Eu5,Eu6,Eu7,Eu8}
allowed to investigate more deeply the interplay of magnetism and superconductivity.
Compared to the past there is now a new aspect to be considered:
not only magnetism competes with superconductivity, but
it can also be involved in the mechanism of
superconductivity itself, as in the case of cuprates, heavy fermions and iron-based superconductors.

The case of the family of iron-based superconductors $EuFe_{2}As_{2}$ \cite{Eu1,Eu2,Eu3,Eu4,Eu5,Eu6,Eu7,Eu8} is particularly
interesting because the ferromagnetic and superconducting transition temperatures are near, where the first is connected
to the $Eu2+$ local magnetic moments. It can also happen that the superconducting critical temperature
is higher than that of the magnetic ordering \cite{Eu1,Eu2,Eu3,Eu4,Eu5,Eu6,Eu7,Eu8}.
In these systems a complex phenomenology of magnetic phases is observed: below the critical superconducting temperature two distinct magnetic transitions take place, the ordering at higher temperature
is associated with the antiferromagnetic interlayer
coupling, whereas the behaviour at lower temperature
might be identified as the change over to a spin-glass state, where
the moments between the layers are decoupled \cite{Eu2,Eu7}.
Usually the spin-glass state \cite{spinglass} occurs in substitutionally
disordered alloys \cite{phon,phon1,phon2}, where, by means of the
long-range Rudermann-Kittel-Kasuya-Yosida
interaction, mediated by conduction electrons, the localized magnetic moments,
randomly distributed, interact. Because not all
magnetic moments can be simultaneously satisfied in
their spin orientation with respect to the others, it happens that
frustration in the magnetic ordering arises. This fact produces an infinite number
of random configurations degenerate in energy but separated by large energy
barriers. In this way the ground state cannot evolve into
another on the experimental time scale.
A typical freezing temperature $T_{SG}$ is associated with the spin-glass state below which
the spins freeze into one of these random
configurations. The magnetic susceptibility in the spin-glasses shows a cusp at $T_{SG}$,
while nothing happens to specific heat other than a broad maximum around $T_{SG}$,
and no Bragg peaks, which usually are a signal of long-range
magnetic order, is found in neutron scattering experiments.
The correct order parameter for these systems has to be related to the probability
that a spin with a given direction at a finite time will have the same direction in the infinite-time limit.
The frozen nature of the spin-glass state is reflected in this order parameter, but no spatial correlations are present as instead happens
in other magnetic order parameters. Is it possible to reproduce this phenomenology connected with the superconducting state inside a theory?
In this paper I will discuss the predictions of the theory on some physical quantities for a multiband spin-glass non phononic $s\pm$-wave superconductor in the framework of Eliashberg equations and I will take as example the particular case of $EuFe_{2}(As_{1-x}P_{x})_{2}$ \cite{Eu5}.
Of course I do not claim to reproduce the complex experimental phenomenology of this material but simply to have indications on the input parameters to be included in the Eliashberg equations in the hope that one day it will be possible to find a material where only the superconducting state and the spin-glass state appear without any other complications.
The starting point will be the theoretical work of M.J. Nass \cite{Nass1,Nass2,Nass3} and J.P Carbotte \cite{CarbotteSP1,CarbotteSP2,CarbotteSP3,CarbotteSP4} that describe a single band spin-glass s-wave phononic superconductor always in the framework of Eliashberg theory.
\section {The Model}
By introducing the order parameter for the spin-glass state as
$q=\lim_{t\rightarrow +\infty}<\textbf{S}_{i}(t)\cdot \textbf{S}_{i}(0)>$
it is possible to describe mathematically this spin freezing \cite{spinglass}.
This order parameter is proportional to the probability that a given spin that has a particular
direction at $t=0$ will still orientated in that direction an infinite time later.
This situation is quite different from having a order parameter in a ferromagnetic
or antiferromagnetic system which reflect space as well as time correlations.
Although each spin is essentially fixed in direction, in the absence
of a magnetic field, upon averaging over all spins the total spin is zero at all temperatures.
By introducing a probability distribution it is possible to reproduce the randomness of the exchange interaction, and then
average over this distribution. It is necessary to use the replica approach in order to carry out the averaging of the
free energy over this distribution of exchange interactions and succeeded in finding a new order parameter defined
as the configuration average of the equal time spin operators at a given site in different replicas of the system \cite{spinglass}.

In the past papers \cite{Nass1,Nass2,Nass3,CarbotteSP1,CarbotteSP2,CarbotteSP3,CarbotteSP4}, the developed theory is on phononic superconductors where it was also added a contribution of antiferromagnetic spin fluctuactions (dynamic part) and spin-glass (static part). In this case, it is not necessary to introduce the dynamic part which is already being responsible for the mechanism of superconductivity but only the static part which is formally equal to the contribution of magnetic impurities with an additional dependence on temperature.
The antiferromagnetic spin fluctuactions have two components \cite{Nass1,Nass2,Nass3,CarbotteSP1,CarbotteSP2,CarbotteSP3,CarbotteSP4}: a dynamical component responsible of $s\pm$ superconductivity
and a static component responsible to spin-glass behaviour that goes to zero at the spin-glass critical temperature $T_{SG}$.
For $T>T_{SG}$ the static component disappears and the material behaves like a normal $s\pm$ superconductor.
In the old phononic low temperature superconductors the dynamic part is, usually, negligible and pair breaking while, in the multiband iron pnictides superconductors, it is the responsible of superconductivity.
The contribution of the spin-glass phase can be represented, in an approximate way, in Eliashberg equations, by a term ($\Gamma^{M}(T)$) similar to that associated with the presence of magnetic impurities but with a temperature dependence. Precisely the magnetic impurities scattering rate \cite{CarbotteSP1,CarbotteSP2,CarbotteSP3,CarbotteSP4} that mimics the spin-glass state is $\Gamma^{M}(T)=\pi N(0)J^{2}S^{2}[1-(\frac{T}{T_{SG}})^{\beta}]$ where $N(0)$ is the total density of states at the Fermi level, $J$ is a exchange constant, $S$ is the spin of the magnetic element and $\beta$ is a number \cite{CarbotteSP1,CarbotteSP2,CarbotteSP3,CarbotteSP4} that can be $1$ or $2$ depending from the physical characteristic of magnetic element ($Eu$ in this case) and of the host material (the specific iron pnictides). At this moment there are not enough data to understand if $\beta$ is 1 or 2 so I solve the Eliashberg equations in the two cases.
This theory stems from the desire to build a very simple model that still manages to grasp the fundamental physics of a multiband spin-glass superconductor. More sophisticated theories \cite{werner} start from multi-orbital Hubbard models
that produce richer phase diagrams and also triplet superconductivity.
For solving the Eliashberg equations are necessary a lot of input parameters connected with the characteristic of physical system. In the following I will refer to $EuFe_{2}(As_{0.835}P_{0.165})_{2}$ a material \cite{Eu5} of iron pinictides family.
The electronic structure of the compound $EuFe_{2}(As_{0.835}P_{0.165})_{2}$ can be approximately described, in principle, as almost all electrons doped iron-based materials, by a three-band model with two electron bands (indicated in the following as bands 1 and 2) and one hole band (indicated in the following as band 3) \cite{tors1}. In this way the gap of hole band, $\Delta_{3}$, has opposite sign to the gaps residing on the electrons bands $\Delta_{1}$ and $\Delta_{2}$.
This compound is especially fascinating since, despite the proximity of the magnetic and superconducting
phases observed at rather high temperatures, there is just a little variation of their transition temperatures to these two phases \cite{Eu5}. The same happens for the stoichiometric material $RbEuFe_{4}As_{4}$ where the superconductivity and a long range magnetic orders exist independently from each other \cite{Eu9}. In this simple model the effect of spin-glasses are simulated by some functions of temperature $\Gamma^{M}\ped{jk}(T)$ that go to zero before $T_{c}$ (precisely to $T_{SG}<T_{c}$) and in this way they do not affect the critical temperature but change the behaviour of some physical quantities below $T_{c}$.

In the iron pnictides the phonons are responsible for the intraband coupling (\textit{ph}) \cite{Mazin_spm} and usually are neglected while
the antiferromagnetic spin fluctuations (\textit{sf}) are connected to interband coupling between holes and electrons bands ($s\pm$ wave model \cite{Aigor,Mazin_spm}).
With the intention to reduce the number of free parameters I use an effective two-band model (band 1 electrons, band 2 holes) where it is not possible to set to zero the intraband coupling and where the electron-boson coupling constants have not an immediate interpretation \cite{dolgoveff,tors2} because this model simulates the true physical situation (three bands) with effective values of electron boson coupling constants in a two bands model.
I investigate what happens in a multiband system and, for simplicity, I study a two bands system that simulates a real three bands system.
In the following the s$\pm$ wave two bands Eliashberg equations \cite{Eliashberg,Chubukov,tors3} are written.
To calculate the critical temperature and the gaps, it is necessary to solve 4 coupled equations: $2$ for the renormalization functions $Z_{j}(i\omega_{n})$ and $2$ for the gaps $\Delta_{j}(i\omega_{n})$, where $j,k$ are band index (that range between $1$ and $2$) and $\omega_{n}$ are the Matsubara frequencies. The imaginary-axis equations \cite{Umma1,Umma2,Umma3} read:
\begin{eqnarray}
&&\omega_{n}Z_{j}(i\omega_{n})=\omega_{n}+ \pi T\sum_{m,k}\Lambda^{Z}_{jk}(i\omega_{n},i\omega_{m})N^{Z}_{k}(i\omega_{m})+\nonumber\\
&&+\sum_{k}\big[\Gamma^{N}\ped{jk}+\Gamma^{M}\ped{jk}(T)\big]N^{Z}_{k}(i\omega_{n})
\label{eq:EE1}
\end{eqnarray}
\begin{eqnarray}
&&Z_{j}(i\omega_{n})\Delta_{j}(i\omega_{n})=\pi
T\sum_{m,k}\big[\Lambda^{\Delta}_{jk}(i\omega_{n},i\omega_{m})-\mu^{*}_{jk}(\omega_{c})\big]\times\nonumber\\
&&\times\Theta(\omega_{c}-|\omega_{m}|)N^{\Delta}_{k}(i\omega_{m})
+\sum_{k}[\Gamma^{N}\ped{jk}-\Gamma^{M}\ped{jk}(T)]N^{\Delta}_{k}(i\omega_{n})\phantom{aaaaaa}
 \label{eq:EE2}
\end{eqnarray}
where $\Gamma^{N}\ped{jk}$ and $\Gamma^{M}\ped{jk}(T)$ are the scattering rates from non-magnetic and magnetic impurities that, in this model, represent the term connected with the spin-glass phase.
For spin-glass superconductors the magnetic impurities scattering rates are $\Gamma^{M}\ped{jk}(T)=c_{jk}\pi N(0)J^{2}S^{2}[1-(\frac{T}{T_{SG}})^{\beta}]=k_{jk}[1-(\frac{T}{T_{SG}})^{\beta}]$
where $c_{jk}$ are weight connected with the bands ($\frac{c_{jk}}{c_{kj}}=\frac{N_{k}(0)}{N_{j}(0)}$ as the usual impurity scattering rates \cite{Umma1,Umma2,Umma3}) and, of course, $k_{jk}=c_{jk}\pi N(0)J^{2}S^{2}$.
I put the non magnetic scattering rates $\Gamma^{N}\ped{jk}$ equal to zero because I suppose to have good single crystals (no disorder).
In the previous equations I have $\Lambda^{Z}_{jk}(i\omega_{n},i\omega_{m})=\Lambda^{ph}_{jk}(i\omega_{n},i\omega_{m})+\Lambda^{sf}_{jk}(i\omega_{n},i\omega_{m})$ and
$\Lambda^{\Delta}_{jk}(i\omega_{n},i\omega_{m})=\Lambda^{ph}_{jk}(i\omega_{n},i\omega_{m})-\Lambda^{sf}_{jk}(i\omega_{n},i\omega_{m})$
where
\[\Lambda^{ph,sf}_{jk}(i\omega_{n},i\omega_{m})=2
\int_{0}^{+\infty}d\Omega \Omega
\alpha^{2}_{jk}F^{ph,sf}_{jk}(\Omega)/[(\omega_{n}-\omega_{m})^{2}+\Omega^{2}], \]
$\Theta$ is the Heaviside function and $\omega_{c}$ is a cutoff
energy.  The quantities $\mu^{*}_{jk}(\omega\ped{c})$ are the elements of the $2\times 2$
Coulomb pseudopotential matrix and finally,
$N^{\Delta}_{k}(i\omega_{m})=\Delta_{k}(i\omega_{m})/
{\sqrt{\omega^{2}_{m}+\Delta^{2}_{k}(i\omega_{m})}}$ and
$N^{Z}_{k}(i\omega_{m})=\omega_{m}/{\sqrt{\omega^{2}_{m}+\Delta^{2}_{k}(i\omega_{m})}}$.
The electron-boson coupling constants are defined as $\lambda^{ph,sf}_{jk}=2\int_{0}^{+\infty}d\Omega\frac{\alpha^{2}_{jk}F^{ph,sf}_{jk}(\Omega)}{\Omega}$.

In order to have the smallest number of free parameter and the simplest model that still grasps the physics of this system, I make further assumptions that have been shown to be valid for iron pnictides \cite{Umma3,Umma1,Umma2}. I assume, following ref. \cite{Mazin_spm} that the total electron-phonon coupling constant is small (the upper limit of the phonon coupling in the usual iron-arsenide compounds is  $\approx0.35$ \cite{Boeri2})so I put, in first approximation, the phonon contribution equal to zero ($\lambda^{ph}_{jk}=0$) and as, following Mazin \cite{Mazincoulomb}, the Coulomb pseudopotential matrix: $\mu^{*}_{jj}(\omega\ped{c})=\mu^{*}_{jk}(\omega\ped{c})=0$
\cite{Umma3,Umma1,Umma2,Mazincoulomb}.
After all these approximations, I write the electron-boson coupling constant matrix $\lambda_{jk}$ in this way:
\cite{Umma3,Umma1,Umma2,Mazin_PhysC_SI}:
\begin{equation}
\vspace{2mm} %
\lambda_{jk}= \left (
\begin{array}{ccc}
          \lambda^{sf}_{11}                  &               \lambda^{sf}_{12}            \\
  &               \lambda^{sf}_{21}=\lambda^{sf}_{12}\nu_{12}               &                 \lambda^{sf}_{22}            \\

\end{array}
\right ) \label{eq:matrix}
\end{equation}
where $\nu_{12}=N_{1}(0)/N_{2}(0)$, and $N_{j}(0)$ is the normal
density of states at the Fermi level for the $j$-th band.
Based on experimental data and theoretical calculations \cite{Umma3,Umma1,Umma2} I choose for the electron-antiferromagnetic spin fluctuation spectral functions $\alpha^2_{jk}F^{sf}_{jk}(\Omega)$ a Lorentzian shape, i.e.:
\begin{equation}
\alpha_{jk}^2F^{sf}_{jk}(\Omega)= C_{jk}\big\{\frac{1}{(\Omega+\Omega_{jk})^2+Y_{jk}^2}-
\frac{1}{(\Omega-\Omega_{jk})^2+Y_{jk}^2}\big\},
\end{equation}
where $C_{jk}$ are normalization constants, necessary to obtain the proper values of $\lambda^{sf}_{jk}$, while $\Omega_{jk}$ and $Y_{jk}$ are the peak energies and the half-widths of the Lorentzian functions, respectively \cite{Umma3}. Following the experimental data \cite{Inosov} I put $\Omega_{jk}=\Omega_{0}$, i.e. I assume that the characteristic energy of spin fluctuations is a single quantity for all the coupling channels, and  $Y_{jk}= \Omega_{0}/2$. The spectral function used here, normalized to one, is shown in the inset (a) of Fig 1.

The factors $\nu_{jk}$ that enter in the definition of $\lambda_{jk}$ (eq. 3) are unknown so I assume that they are equal, for example, to the $Ba(Fe_{1-x}Rh_{x})_{2}As_{2}$ electron doped case \cite{ghigo} so $\nu_{12}=0.8333$ as well as the coupling constants \cite{ghigo} and I change lightly just the value of $\lambda_{22}$ for obtaining the correct critical temperature $T_{c}=22$ K. At the end the values are $\lambda_{11}=1.00$, $\lambda_{12}=-0.17$ and $\lambda_{22}=2.65$ for a averaged coupling constant $\lambda_{t}=\frac{\Sigma_{jk}N_{j}(0)\lambda_{jk}}{\Sigma_{j}N_{j}(0)}=1.75$.
For iron pnictides it was experimentally found \cite{Paglione, Inosov1} that the empirical law $\Omega_{0}=2T_{c}/5$ holds, therefore the value of the energy peak $\Omega_{0}$ of the Eliashberg spectral functions $\alpha^2_{jk}F_{jk}^{sf}(\Omega)$ is fixed. To finish, in the numerical calculations I used a cut-off energy $\omega_{c}=180$ meV.
These input parameters produce, by numerically solving the Eliashberg equations, exactly a critical temperature of $22$ K.
%
\section{Calculation of superconducting gaps}
In the iron pnictides, usually, the impurities are almost all concentrated in one band: i.e. in the hole band for the electron doped materials as this case and in the electron band \cite{ummares,tors4} for the hole doped materials \cite{maxres}. This means that, in the electrons doped materials, $k_{22}>>k_{11},k_{12}$. I choose $k_{11}=k_{12}=0.2 k_{22}$ as happen in the $Ba(Fe_{1-x}Co_{x})_{2}As_{2}$ \cite{ummares}.
By using the typical parameters of iron pnictides and spin-glass systems I find that $k_{22}\simeq 3.1$ meV ($N(0)=5.6$ states/eV, $S=7/2$, $J=0.12$ meV and $T_{SG}=15$ K) \cite{Eu5,sados}. Because the true values of the parameters in the last bracket are just approximative I solve the Eliasberg equations for values close to 3.1 as $k_{22}=0, 1, 2, 3, 3.5, 4, 4.05, 5$ meV in the two cases: $\beta=1$ and $\beta=2$. In the ideal case it would be necessary to know the law that links $T_{SG}$ at the value of $k_{22}$. Here $T_{SG}=15$ K is an experimental input \cite{Eu5}.
In the Figs 1 and 2 the temperature dependence of gaps $\Delta_{1,2}(i\omega_{n=0})$ are shown. It is possible to see in Fig 1 that the absolute values of the gaps with increasing temperature at first increases until $T_{SG}$, and then decreases.
This behaviour appears when magnetic impurities (also without temperature dependence) and disorder are present, both in $s++$ and $s\pm$ superconductor or if only disorder is present in $s\pm$ two bands superconductor \cite{gammagap}.

For $k_{22}=5 $ meV and $\beta=1$, reentrant superconductivity is obtained. As it is shown in Fig 2 in the case $\beta=1$  the effect is similar but stronger and, besides having reentrant superconductivity for $k_{22}=5 $ meV, it is possible to see an even more complex situation for $k_{22}=4.05 $ meV. In the last case the system has three different critical temperatures: I think that it would be difficult to observe this behaviour in a real system because it arises from a fine tuning of the input parameter. In a s-wave superconductor the magnetic order destroys the superconductivity but the temperature weakens both the magnetic order and the coupling between the electrons in the Cooper pairs so the "reentrant" behaviour can emerge from the balance between the effect of magnetism and temperature.
I solved the Eliashberg equations, for completeness, also in the case $k_{12}=0.2 k_{22}$ and $k_{11}=k_{22}$ always with $\beta=1$ and $\beta=2$. The results are shown in Fig 3 and are similar to previous ones in the general trend as a function of the value of $k_{22}$. In all case, of course, for $T>T_{SG}$ the effect of "magnetic impurities" disappeared and the behaviour is the same of a standard two bands superconductor.
\begin{figure}[H]
\begin{center}
\includegraphics[keepaspectratio, width=\columnwidth]{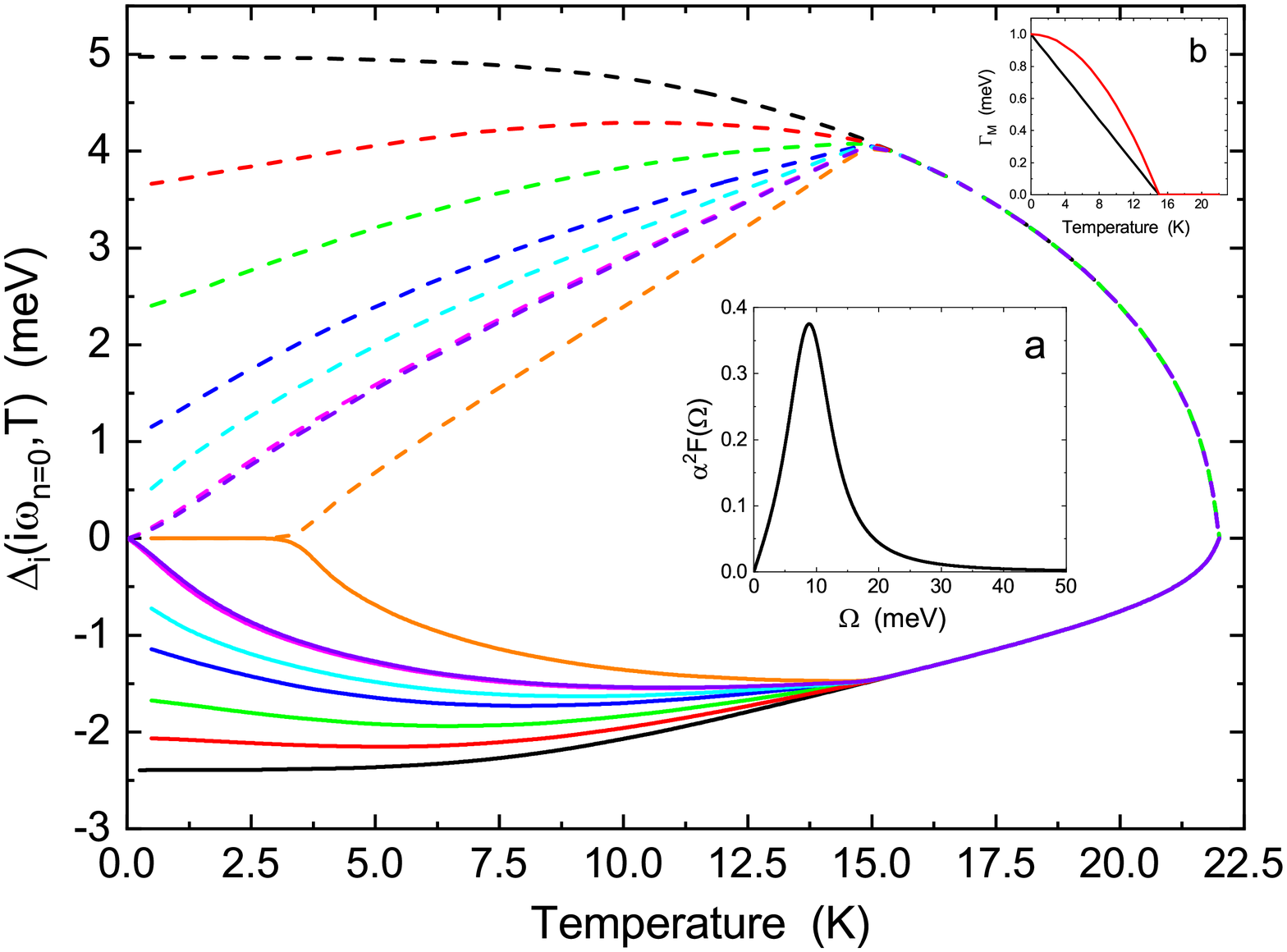}
\caption{(Color online) The gaps $\Delta_{i}(i\omega_{n=0})$ in function of temperature obtained by solving the Eliashberg equations on imaginary axis: solid lines for $\Delta_{1}(i\omega_{n=0})$, dashed lines for $\Delta_{2}(i\omega_{n=0})$ in the case $k_{11}=k_{12}=0.2k_{22}$ and $\beta=1$. Black lines for $k_{22}=0$ meV, red lines for $k_{22}=1$ meV, green lines for $k_{22}=2$ meV, dark blue lines for $k_{22}=3$ meV, cyan line for $k_{22}=3.5$ meV, magenta lines for $k_{22}=4$ meV, violet lines for $k_{22}=4.05$ meV and orange lines for $k_{22}=5$ meV. In the inset (a) the antiferromagnetic spin fluctuactions function, normalized to one, is shown while in the inset (b) the temperature dependence of $k_{jk}$ is shown (black solid line $\beta=1$ and red solid line $\beta=2$) with $k_{jk}=1$ .
}
\end{center}
\end{figure}

\begin{figure}[H]
\begin{center}
\includegraphics[keepaspectratio, width=\columnwidth]{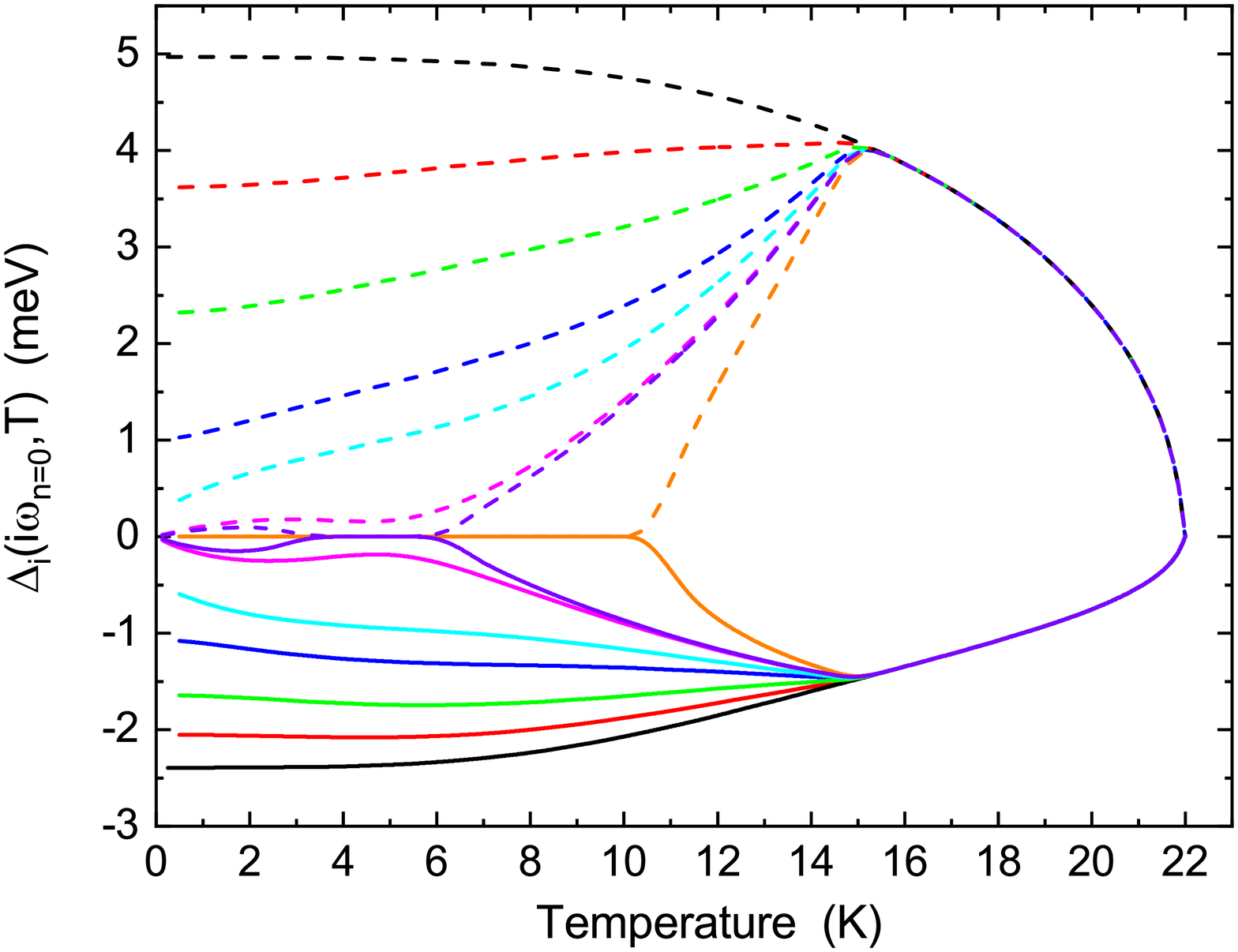}
\caption{(Color online) The gaps $\Delta_{i}(i\omega_{n=0})$ in function of temperature obtained by solving the Eliashberg equations on imaginary axis: solid lines for $\Delta_{1}(i\omega_{n=0})$, dashed lines for $\Delta_{2}(i\omega_{n=0})$ in the case $k_{11}=k_{12}=0.2k_{22}$ and $\beta=2$. Black lines for $k_{22}=0$ meV, red lines for $k_{22}=1$ meV, green lines for $k_{22}=2$ meV, dark blue lines for $k_{22}=3$ meV, cyan line for $k_{22}=3.5$ meV, magenta lines for $k_{22}=4$ meV, violet lines for $k_{22}=4.05$ meV and orange lines for $k_{22}=5$ meV.}
\end{center}
\end{figure}

\begin{figure}[H]
\begin{center}
\includegraphics[keepaspectratio, width=\columnwidth]{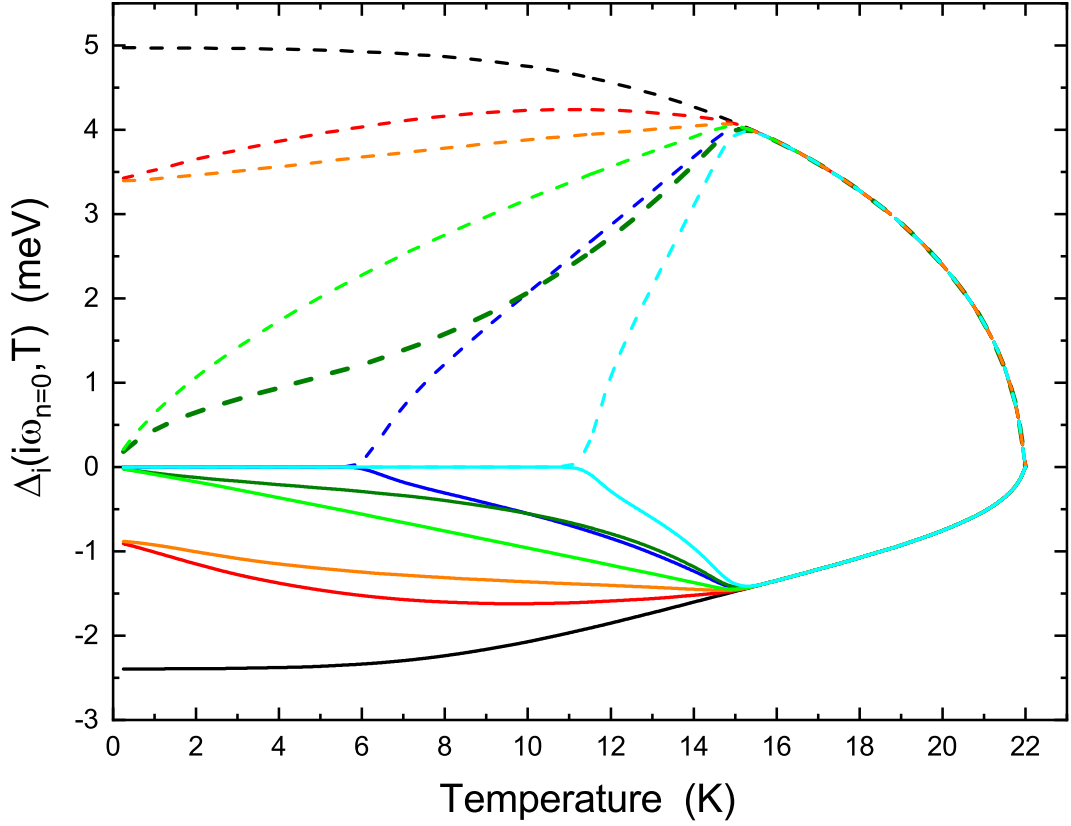}
\caption{(Color online) The gaps $\Delta_{i}(i\omega_{n=0})$ in function of temperature obtained by solving the Eliashberg equations on imaginary axis: solid lines for $\Delta_{1}(i\omega_{n=0})$, dashed lines for $\Delta_{2}(i\omega_{n=0})$ in the case $k_{11}=k_{22}=5k_{12}$. Black lines for $k_{22}=0$ meV, red lines for $k_{22}=1$ meV and $\beta=1$, orange lines for $k_{22}=1$ meV and $\beta=2$, green lines for $k_{22}=3$ meV and $\beta=1$, olive lines for $k_{22}=3$ meV and $\beta=2$, dark blue lines for $k_{22}=5$ meV and $\beta=1$ and magneta lines for $k_{22}=5$ meV and $\beta=2$.}
\end{center}
\end{figure}
\section{Calculation of the penetration depth}
 The penetration depth (or the superfluid density as it is shown in Figs. 4, 5 and 6) can be computed starting from the renormalization functions $Z_{j}(i\omega_{n})$ and the gaps $\Delta_{j}(i\omega_{n})$ by using the following formula: \cite{lambda}
\begin{eqnarray}
\lambda^{-2}(T)=(\frac{\omega_{p}}{c})^{2} \sum_{j=1}^{2}w_{j}\pi T \sum_{n=-\infty}^{+\infty}\frac{\Delta_{j}^{2}(\omega_{n})Z_{j}^{2}(\omega_{n})}{[\omega^{2}_{n}Z_{j}^{2}(\omega_{n})+\Delta_{j}^{2}(\omega_{n})Z_{j}^{2}(\omega_{n})]^{3/2}}\label{eq.lambda}
\end{eqnarray}
where $\omega_{p,i}$ is the plasma frequency of the $i$-th band and $\omega_{p}$ is the total plasma frequency in order that the $w_{j}=\left(\omega_{p,j}/\omega_{p}\right)^{2}$ are the weights of the single bands.

The low-temperature value of the penetration depth $\lambda_L(0)$ should, in principle, be related to the plasma frequency by $\omega_p=c/\lambda_L(0)$ \cite{tors5} and appears as a multiplicative factor of the summation. Here $w_{1}=0.72$ and $w_{2}=0.28$ as in the Co doped iron compounds \cite{ummares}. In principle, it is important the calculation of superfluid density (penetration depth) in order to compare theoretical predictions with experiment because it is easier to find these measurements in literature \cite{tors6}. In the Figs 4, 5 the superfluid density in function of temperature is shown when $k_{11}=k_{12}=0.2 k_{22}$, $k_{22}=0, 1, 2, 3, 3.5, 4, 4.05, 5$ meV with $\beta=1$ and $\beta=2$. In Fig. 6 I show the superfluid density when $k_{12}=0.2 k_{11}=0.2 k_{22}$, $k_{22}=0, 1, 3, 5$  meV with $\beta=1$ and $\beta=2$. These results are a clear prediction of possible situations that can be easily identified. Unfortunately, there is still no experimental data to compare with these theoretical predictions. The behavior of the penetration depth as a function of temperature shows how the presence of a spin-glass state in competition with superconductivity substantially changes the phase diagram of a superconductor making it extremely richer.
Also for the superfluid density it is clear when reentrant superconductivity appears as well as the case of three critical temperatures (see Fig 5 violet line).
In the Figs 1 and 2 in the cases with $k_{22}=4.00$ meV and $k_{22}=4.05$ meV the values of $\Delta_{1}(i\omega_{n=0})$ and $\Delta_{2}(i\omega_{n=0})$, at very low temperatures, are almost zero but the corresponding superfluid density at the same temperatures
is different from zero in an appreciable way. As is it possible? From the numerical solution of Eliashberg equation in the standard case 
(when the $k_{jk}$ are equal to zero) the maximum value of $|\Delta_{i}(i\omega_{n})|$ is for $n=0$ and the $|\Delta_{i}(i\omega_{n})|$ decreases when $|n|$ increases while, when the $k_{jk}$ are different from zero, the dependence from $|n|$ is different and not usual. In the inset of Fig. 5 the calculated values of $\Delta_{1}(i\omega_{n})$ and $\Delta_{2}(i\omega_{n})$, in the $k_{22}=4.05$ meV and $\beta=2$ case at $T=0.125$ K, in function of $n$ is shown. In this case it is possible to see that the dependence of $\Delta_{j}(i\omega_{n})$ from $n$ is not standard. In this case the maximum value of $|\Delta_{i}(i\omega_{n})|$ is not more for $n=0$ so also if 
$\Delta_{i}(i\omega_{n=0})\simeq 0$ meV the corresponding superfluid density can be different from zero.
\begin{figure}[H]
\begin{center}
\includegraphics[keepaspectratio, width=\columnwidth]{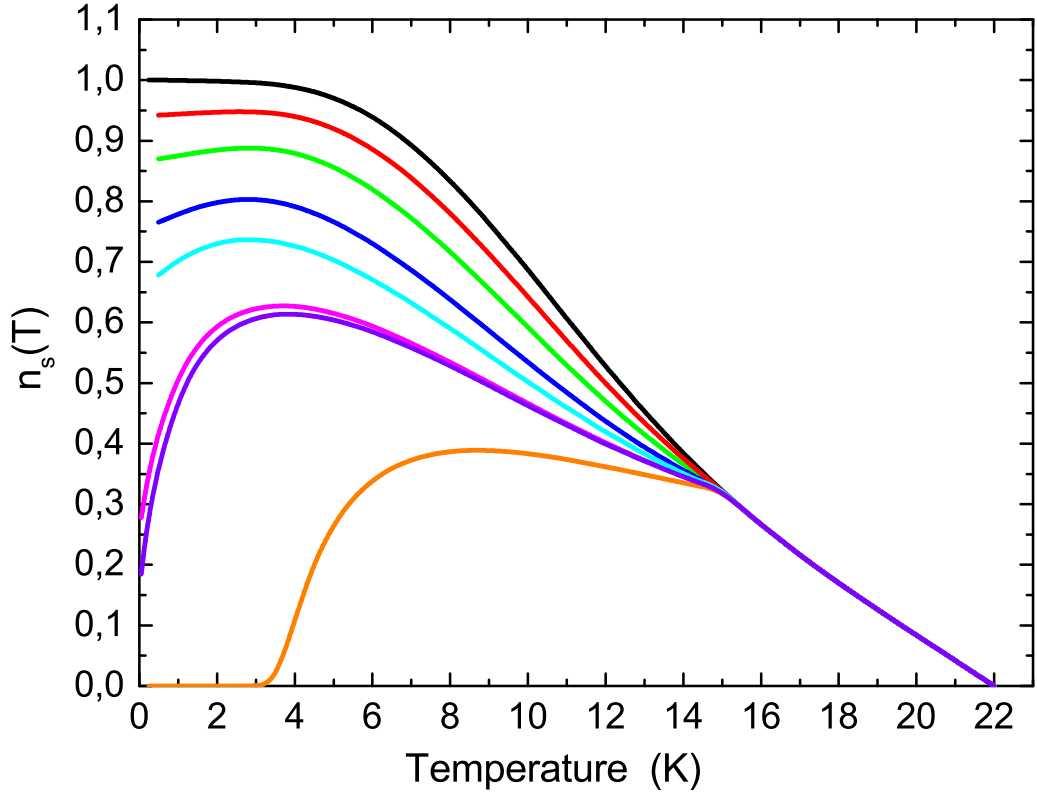}
\caption{(Color online) The superfluid density $n_{s}(T)$, normalized at the value at T=0 K in the case $k_{22}=0$, in function of temperature, obtained by solving the Eliashberg equations on imaginary axis in the case $k_{11}=k_{12}=0.2k_{22}$ and $\beta=1$. Black line for $k_{22}=0$ meV, red line for $k_{22}=1$ meV, green line for $k_{22}=2$ meV, dark blue line for $k_{22}=3$ meV, cyan line for $k_{22}=3.5$ meV, magenta lines for $k_{22}=4$ meV, violet lines for $k_{22}=4.05$ meV and orange line for $k_{22}=5$ meV.}
\end{center}
\end{figure}

\begin{figure}[H]
\begin{center}
\includegraphics[keepaspectratio, width=\columnwidth]{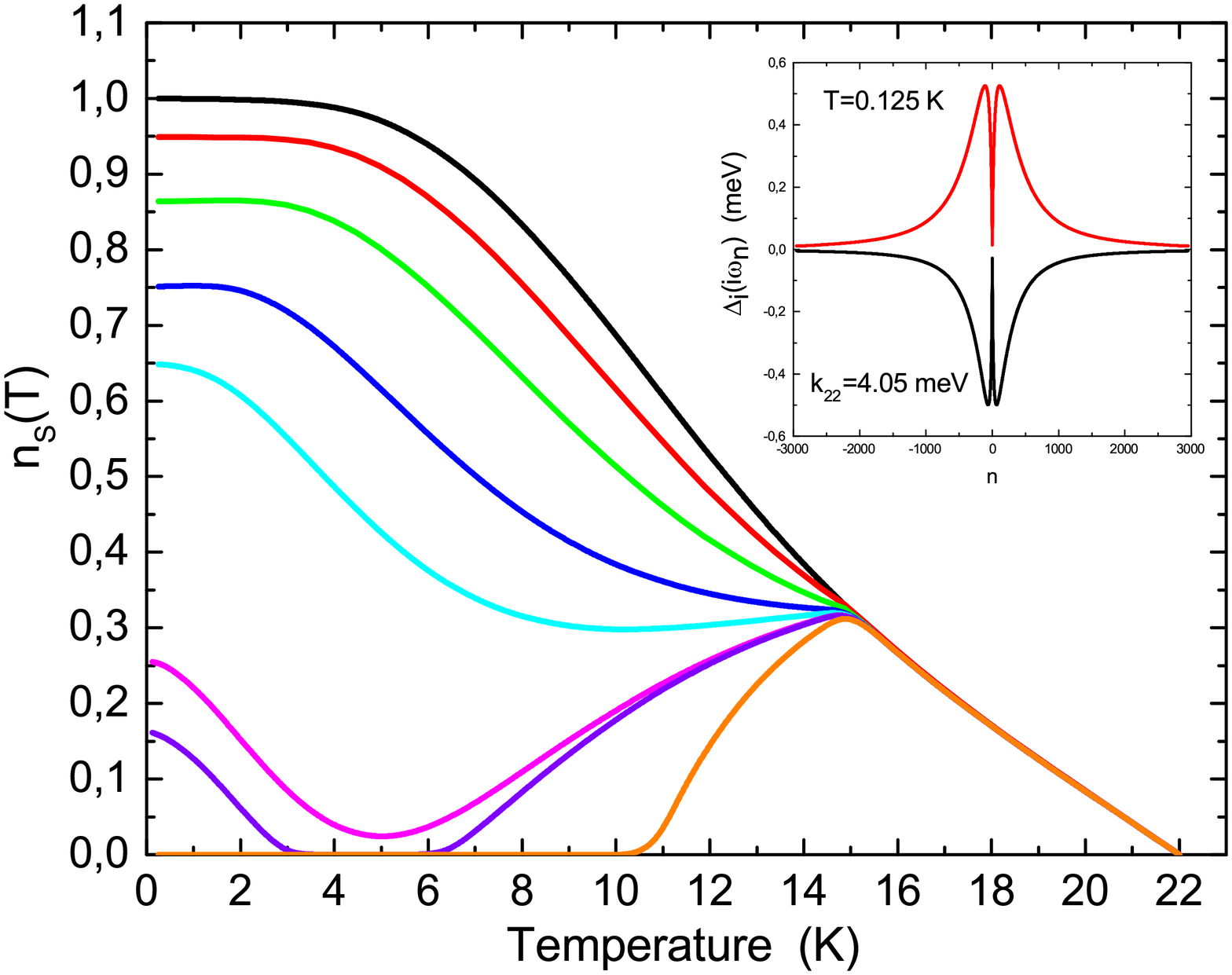}
\caption{(Color online) The superfluid density $n_{s}(T)$, normalized at the value at T=0 K in the case $k_{22}=0$, in function of temperature, obtained by solving the Eliashberg equations on imaginary axis in the case $k_{11}=k_{12}=0.2k_{22}$ and $\beta=2$. Black line for $k_{22}=0$ meV, red line for $k_{22}=1$ meV, green line for $k_{22}=2$ meV, dark blue line for $k_{22}=3$ meV, cyan line for $k_{22}=3.5$ meV, magenta line for $k_{22}=4$ meV, violet lines for $k_{22}=4.05$ meV and orange line for $k_{22}=5$ meV. In the inset, the dependence, obtained by numerical solution of Eliashberg equations in the $k_{22}=4.05$ meV case at $T=0.125$ K, of the two order parameters $\Delta_{j}(i\omega_{n})$ from the index $n$ is shown.}
\end{center}
\end{figure}

\begin{figure}[H]
\begin{center}
\includegraphics[keepaspectratio, width=\columnwidth]{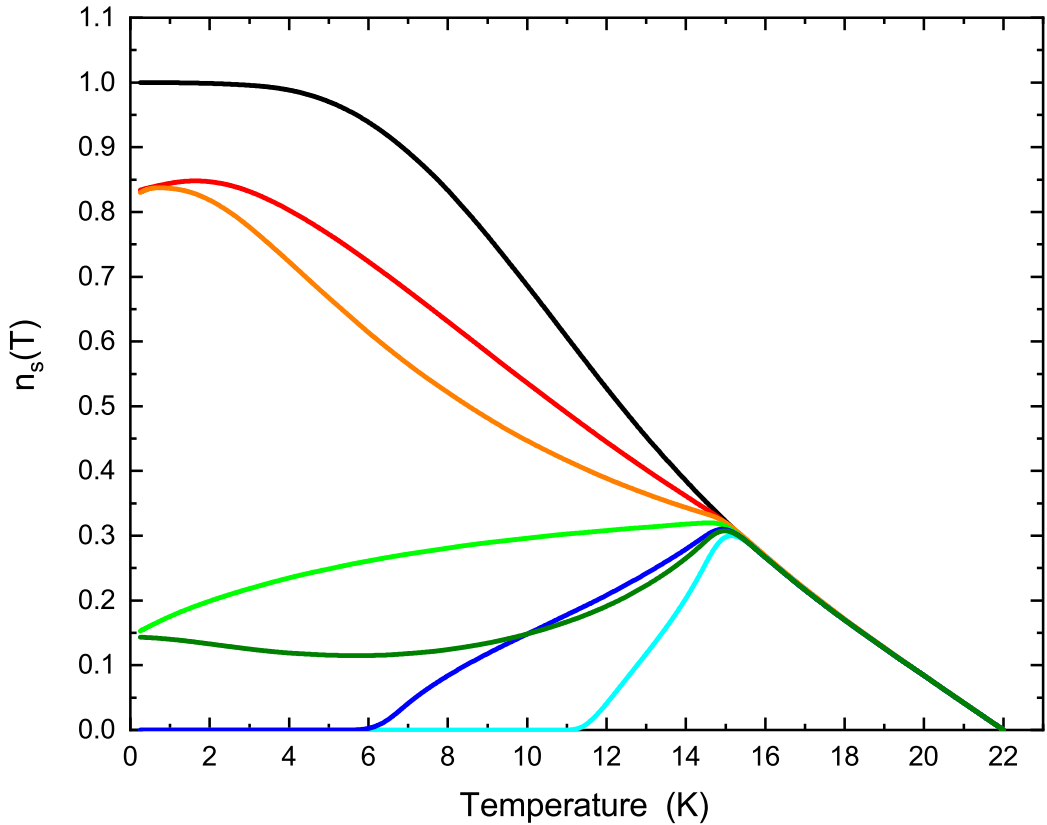}
\caption{(Color online) The superfluid density $n_{s}(T)$, normalized at the value at T=0 K in the case $k_{22}=0$, in function of temperature obtained by solving the Eliashberg equations on imaginary axis in the case $k_{11}=k_{22}=5k_{12}$. Black line for $k_{22}=0$ meV, red line for $k_{22}=1$ meV and $\beta=1$, orange line for $k_{22}=1$ meV and $\beta=2$, green line for $k_{22}=3$ meV and $\beta=1$, olive line for $k_{22}=3$ meV and $\beta=2$, dark blue line for $k_{22}=5$ meV and $\beta=1$ and magenta line for $k_{22}=5$ meV and $\beta=2$.}
\end{center}
\end{figure}
%
\section{Conclusions}
In conclusion, I have calculated the temperature dependence of gaps and superfluid densities for a two bands non phononic $s\pm$-wave spin-glass superconductor. In general, the temperature dependence of superconducting properties shows a lot of different behaviours that should be observable in experiment. In this system two competing orders modulated by temperature are present.
The magnetic order breaks down the superconductivity, but in this case it also dependent on temperature as well as the superconducting electron-boson coupling. The temperature weakens both but in a different manner so from this competition can born a more complex phase diagram. In addition, reentrant behavior could be a possible signature of a spin-glass state.
\section{ACKNOWLEDGMENTS}
The author acknowledges support from the MEPhI Academic Excellence Project (Contract No. 02.a03.21.0005). \\

%
\end{document}